\documentclass{article}

\usepackage[preprint]{neurips_2024}

\usepackage[utf8]{inputenc} 
\usepackage[T1]{fontenc}    
\usepackage{hyperref}       
\usepackage{url}            
\usepackage{booktabs}       
\usepackage{amsfonts}       
\usepackage{nicefrac}       
\usepackage{microtype}      
\usepackage{xcolor}         
\usepackage{graphicx}

\title{LISA Technical Report: An Agentic Framework for Smart Contract Auditing}

\author{%
  Izaiah Sun, Daniel Tan, Andy Deng\\
  Agent Lisa Team\\
  Singapore \\
}

\begin{document}

\maketitle

\begin{abstract}
We present LISA, an agentic smart contract vulnerability detection framework that combines rule-based and logic-based methods to address a broad spectrum of vulnerabilities in smart contracts. LISA leverages data from historical audit reports to learn the detection experience (without model fine-tuning), enabling it to generalize learned patterns to unseen projects and evolving threat profiles. In our evaluation, LISA significantly outperforms both LLM-based approaches and traditional static analysis tools, achieving superior coverage of vulnerability types and higher detection accuracy. Our results suggest that LISA offers a compelling solution for industry: delivering more reliable and comprehensive vulnerability detection while reducing the dependence on manual effort.
\end{abstract}

\section{Introduction}

Smart contracts have become integral across blockchain ecosystems, powering decentralized finance (DeFi), nonfungible tokens (NFTs), governance mechanisms, and many other applications. Their appeal lies in their transparency, automation, and the guarantee that the code, once deployed, executes exactly as written. However, this also means that vulnerabilities in smart contracts carry serious risks: bugs or logic errors may result in irrevocable financial loss, degraded trust, or exposure to attack. Because smart contracts are immutable once deployed, any flaw that is exploited typically cannot be patched simply by redeploying, making prevention, detection, and rigorous auditing essential.

Decentralized Finance (DeFi) continues to present a rapidly growing target surface for financial loss driven by smart contract vulnerabilities and logic flaws. According to recent reports, losses from DeFi exploits remain substantial: for example, DeFi protocols lost more than US\$1.029 billion in 2024 from 339 reported incidents, which accounted for more than 80\% of the losses in blockchain security incidents in that period~\cite{slowmistReport}. Between 2020 and 2023, cumulative losses from DeFi exploits have been estimated at approximately US\$58.78 billion~\cite{thedefiantReport}. Meanwhile, even in 2024 alone, hackers stole approximately US\$2.2 billion in cryptocurrency with DeFi platforms among the top affected entities~\cite{ChainalysisReport, CryptoCrimeReport}. These figures underscore that vulnerabilities in smart contracts, both in classical rule-based categories and in logical / business-logic flaws~\cite{Web3Bugs}, are not just academic concerns but have major real-world financial stakes.

Existing vulnerability detection tools and approaches fall broadly into two categories:
\begin{enumerate}
    \item Rule-based / static analysis methods: These rely on manually curated patterns or rules (e.g. detecting reentrancy, integer overflow, unbounded loops, misuse of external calls). They are effective for known vulnerabilities, generally efficient, and relatively lightweight. However, they often fail to detect logic vulnerabilities, subtle business logic flaws, or vulnerabilities that arise from combinations of patterns~\cite{Web3Bugs}. They may generate many false negatives under unusual code structures or when new vulnerability types emerge that do not match any existing rule. Meanwhile, false positives of these tools are not acceptable. with less than 10\% precision for popular open source tools~\cite{Kaixuan2024SAST}.
    \item AI- or large language model (LLM)- driven reasoning approaches: These can generalize better to previously unseen code, understand context, and sometimes detect logic flaws. On the flip side, they can suffer from false positives~\cite{david2023needmanualsmartcontract}, hallucinations or overgeneralization, or lack of interpretability, even for models fine-tuning on smart contract bugs~\cite{Ma2025Combining}. 
\end{enumerate}

A promising but under-utilized source of insight is the body of historical smart contract audit reports. 
These reports include real vulnerability instances, including aspects of logic and business context, together with human reasoning, diagnostics, and often the context of the surrounding code. 
In addition, the audit reports are keeping updating that new vulnerabilities will be included in the dataset.
Although many tools are aware of audit reports, such as PropertyGPT~\cite{liu2024propertygpt}, few take advantage of them in a way that generalizes across projects without exposing proprietary details or overfitting to specific contract patterns.

To fill these gaps, we introduce LISA, an agentic framework for smart contract vulnerability detection that is based on learning from the latest audit reports and attack events. 
LISA is a pure LLM-powered framework designed for both rule-based and logic-based vulnerabilities.
Crucially, LISA learns from historical audit reports, not by fine-tuning large models in a data-sensitive way, but by extracting and internalizing detection experience, patterns, and reasoning heuristics from audit history. This design allows LISA to generalize effectively to new projects, including unseen contract styles or libraries, and to adapt to evolving threat patterns without exposing proprietary model internals.

In our empirical evaluation, LISA is compared with state-of-the-art LLM-based detection systems and leading static analysis tools. 
The results show that LISA achieves notably higher coverage of vulnerability types (especially logic vulnerabilities beyond simple rule patterns) and superior accuracy (i.e., higher precision and recall), reducing false positives and false negatives relative to the baseline methods. 
These benefits translate into lower manual audit effort, stronger security guarantees for deployed contracts, and improved developer confidence.

\section{Scope and Limitation of this Technical Report}

In this technical report, we focus on real-world smart contract auditing scenarios, particularly those arising from professional auditing contests and competitive audit platforms such as Code4Rena~\cite{Code4Rena}, Secure3~\cite{Secure3} and real-world attack events. We ground our evaluation and threat modeling in audits of actual source code in deployed or preparing-for-deployment projects, rather than synthetic or academic datasets.
We consider these settings especially valuable because they reflect the complexity, diversity, and unpredictability of real smart contract code, including dependencies, style heterogeneity, and logic that may not conform to patterns seen in clean academic datasets.

A key limitation of this technical report lies in the coverage of vulnerability types in the dataset used for evaluation. In many real-world audits, especially on contest platforms like Code4rena or Secure3, contracts have already been subject to static analysis tools and manual review, and simple, well-known flaws (such as integer overflow/underflow, basic reentrancy, boundary arithmetic errors) are often already detected and patched before or during the audit process. As a result, our evaluation is biased towards logic vulnerabilities and more complex traditional vulnerabilities (for example, read-only reentrancy, subtle state transition logic, inter-contract dependency flaws) rather than simple pattern-based bugs. Consequently, the results presented in this report do not reflect performance on trivial classes of bugs, and the detection rates for those simple vulnerability types may appear lower or omitted in the metrics, not because LISA is weak there, but because those cases are underrepresented in the audit datasets we use.

\section{Overall Design of LISA}

LISA is built on a modular, agentic framework composed of three main components: a Knowledge Base, a Scheduler, and multiple Detection Agents (both specialized and a fallback general agent). This design enables LISA to leverage historical audit experience, dynamically allocate detection efforts, and handle a broad spectrum of vulnerability types in real-world smart contract source code.

\begin{figure}
    \centering
    \includegraphics[width=\linewidth]{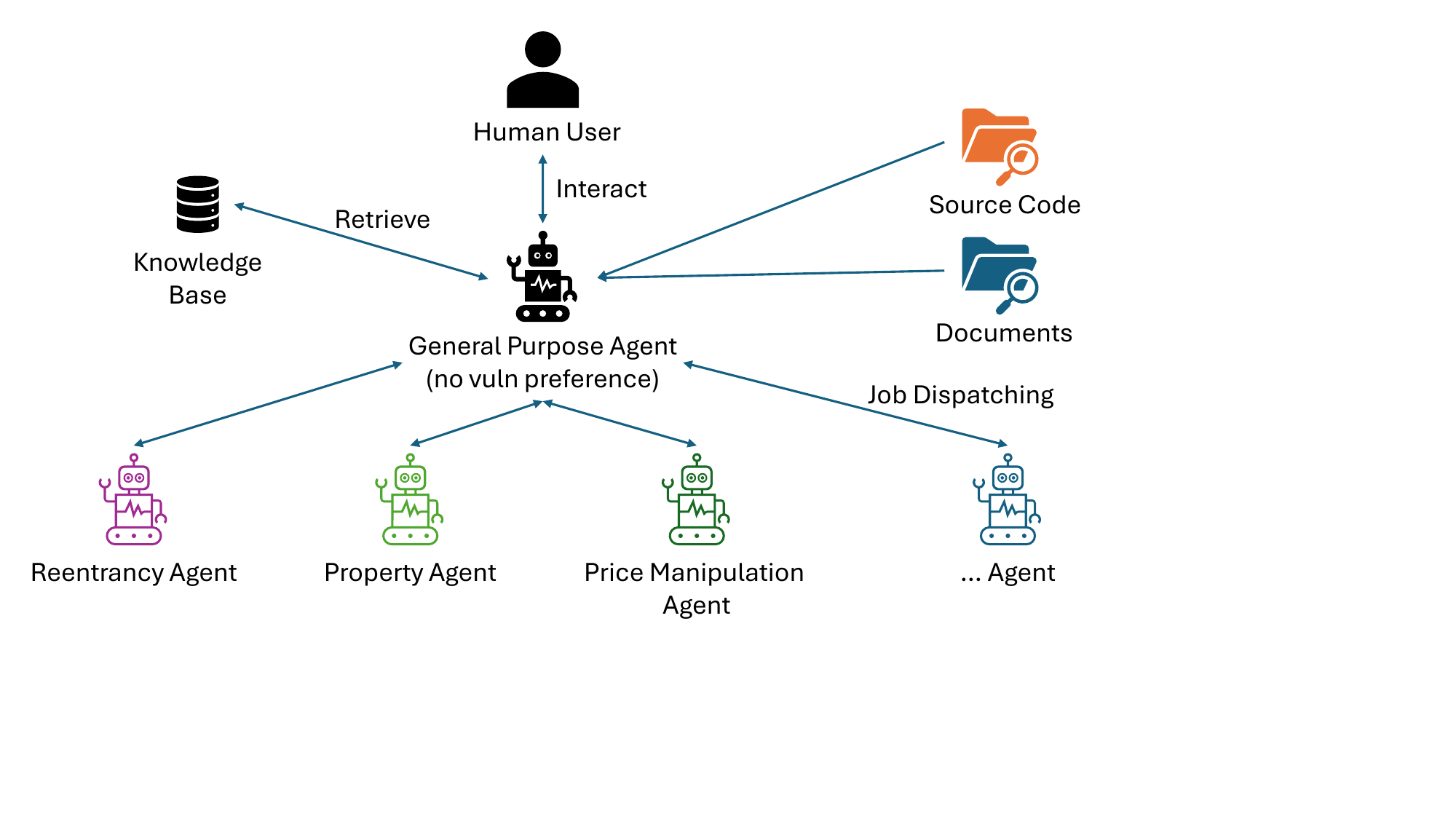}
    \caption{Overall Design of LISA}
    \label{fig:overall}
\end{figure}

\subsection{Knowledge Base}
The Knowledge Base is the foundational component of LISA. 
It captures structured information extracted from real-world smart contract audit reports (especially from competitive audit platforms like Code4rena), including vulnerability type templates, logic pattern heuristics, severity gradings, example code fragments, and human reasoning annotations. 
It stores both rule-based patterns (e.g., typical control-flow or data-flow markers that correspond to known issues such as reentrancy, unchecked external call returns, access control misconfigurations) and logic-based contexts (for example, state transition invariants, expected invariants across contract modules, business logic expectations). 
The Knowledge Base is maintained via continuous ingestion of new audit findings; when new vulnerabilities are discovered, their relevant code contexts and reasoning traces are abstracted and added (in anonymized / generalized form) to enable pattern matching in future audits. 
The Knowledge Base also supports metadata (e.g. project size, contract complexity, language version) so that LISA can factor context when selecting which agents are likely to be effective. 
Importantly, it does not store proprietary detection algorithms, nor does it use private code un-sanitized; everything is abstracted so that matching is on patterns and context rather than leaking internal logic.

\subsection{Scheduler \& General Purpose Agent}
The Scheduler is the brain of LISA that orchestrates the detection flow. 
When new smart contract source code is submitted for audit, the Scheduler first performs a pre-analysis step to extract syntactic features, module boundaries, external call dependencies, function call graphs, etc. 
It queries the Knowledge Base for which vulnerability types are plausible in this code base (based on similarities with past cases, presence of certain patterns, and contract metadata). Based on this, the Scheduler assigns specialized Detection Agents to relevant code areas. 
For example, if the pre-analysis reveals external calls plus unguarded state updates, the Scheduler may dispatch a logic vulnerability agent; for parts of the code with familiar patterns of reentrancy or overflow, the rule-based agents may be assigned. 
The Scheduler supports parallel execution of agents where appropriate (for example, one for rule-based, another for logic-based) and handles sequencing where the output of some agents may guide or pace others (for example, logic agent might depend on the results of rule-based checks). 
If the Scheduler finds that none of the specialized agents in the current set matches the pattern, either because the code context is novel, or the Knowledge Base has no strong matching templates, it falls back to invoke the General Agent for that portion. 
Additionally, the Scheduler manages priorities: agents dealing with high-severity potential vulnerabilities may get more resources; conflicting agents’ findings are reconciled. Finally, the Scheduler aggregates all agent outputs into a unified finding set and normalizes them (e.g. severity levels, confidence scores, deduplication of overlapping findings).

The General Agent is a part of the scheduler. 
The Scheduler routes to the General Agent in situations where code exhibits no strong matches in the Knowledge Base, or the patterns are ambiguous, novel, or beyond the current coverage of specialized agents. 
The General Agent employs more generic detection strategies: lighter static checks, broad heuristic scanning, simple logic reasoning, possibly code similarity checks, and sanity checks (e.g. ensuring state gets updated after external calls, minimal access control checks, etc.). 
Its analyses are designed to be less deep or narrower in scope than specialized agents, and the confidence or severity scores from the General Agent are correspondingly lower (or flagged for manual review). 
The use of the General Agent ensures coverage, LISA is less likely to miss unknown or emerging types simply because no specialized agent existed, but also means there is a trade-off: findings from the General Agent are more likely to need human verification, may have higher false positive rates, and lower severity precision. In practice, the General Agent acts as an early warning net, ensuring no code component is entirely unexamined. 

\subsection{Domain Specific Agents}
LISA includes a cadre of Specialized Agents, each tailored to one or more specific classes of vulnerabilities. Each Specialist Agent is optimized for detection of vulnerabilities of that class by using strategies drawn from the Knowledge Base, leveraging both rule-based and logic-based detection where applicable:
\begin{enumerate}
    \item Rule-Pattern Agents: These specialize in well-known, relatively deterministic vulnerabilities such as reentrancy under certain control‐flow patterns, unchecked external call return values, integer overflow / underflow (subject to compiler version). They rely heavily on pattern templates from the Knowledge Base, combined with static analysis and heuristics. The LLM in such an agent is helping to reduce false positives.
    \item Logic Vulnerability Agents: These focus on business logic flaws, state transitions, invariants, cross-contract interactions, unintended side-effects, misuse of dependency contracts, or vulnerabilities that arise only when sequence of operations or contract interactions unfolds in a certain way. Because logic vulnerabilities often do not correspond to simple static rules, these agents use more sophisticated reasoning, context inference from audit examples in the Knowledge Base, and may apply symbolic reasoning, dataflow/control flow graphs, or other logic checks.
    \item Project-Specific Agents: Depending on project domain (for example, DeFi lending, tokenomics, access control, upgradable contracts), there may be specialized agents that understand domain conventions, gas usage patterns, external library patterns, or known risky patterns (e.g., cross-chain bridging, oracles). These agents can have very high precision in their domain because they operate under narrower, well-understood classes.
\end{enumerate}
For each type of vulnerability, a saperate agent is designed semi-automatically, both with the knowledge from knowledge base and the confirmation from the human experts.

Each Specialized Agent outputs findings annotated with what pattern or example in the Knowledge Base was matched or what heuristic/rule triggered the detection. They also include confidence estimations (based on how similar the code is to stored examples, how complete the matching is, etc.) and also basic explanation trace so that users (auditors / developers) can see why a vulnerability was flagged.

\subsection{Merging of Detection Results}

Once the various Detection Agents (specialized and general) complete their analyses, LISA performs a multi-stage merging procedure to produce the final report. The goal is to reduce false positives and false negatives, ensure consistency, and surface only those vulnerabilities that are credible and relevant. The merging process consists of three main checks:
\begin{enumerate}
    \item \textbf{Factual Error Checking.} Each candidate vulnerability flagged by an agent is first subjected to a factual validation step to ensure that the detection corresponds to an actual issue in the source code. This involves parsing the relevant code segment(s) to verify that the condition asserted by the agent is indeed present (e.g. function call ordering, external calls, state variable updates). If the agent claims a vulnerability but the required pattern or code behavior does not hold upon closer inspection (for example, the external call is actually guarded, or a required condition is missing), the finding is discarded or downgraded. Agents provide enough trace data (line numbers, AST or CFG node references, or similar metadata) so that LISA can re-verify factual correctness.
    \item \textbf{Knowledge Base Cross-Validation.} After factual correctness, each finding is verified against the entries of the knowledge base. The system looks for matching or similar vulnerability templates, past audit examples, or logic patterns that resemble the current finding. If a candidate vulnerability matches a well-understood template (from past audits) with high similarity, its confidence is bumped up; if there is no matching pattern or only weak similarity, the confidence remains lower or it may be flagged for manual review. This step helps LISA leverage the learned experience and avoid 'hallucinated' or spurious findings that do not match the historical precedent.
    \item \textbf{Project-Level Invariant Checking.} Finally, LISA evaluates whether a detected vulnerability violates one or more project-level invariants, i.e., properties or constraints that are expected to always hold for the particular project under audit. These invariants may be explicit (specified by the project developers or derived from documentation) or inferred (from past project behavior, from similar contracts in the Knowledge Base). Examples include invariants such as 'total supply of a token should never change except through mint / burn functions', 'balances must remain nonnegative', 'state transitions must not allow bypass of access control', or 'contract must have a pause mechanism under emergency conditions'. If a finding implies a violation of one of these project invariants, its severity or priority is raised. Conversely, if the finding does not contradict any such invariant or the invariant is weak/not applicable, the finding is assigned a lower risk ranking.
\end{enumerate}

\section{Evaluation}

To compare the performance of LISA on real-world projects. The evaluation is conducted in three parts. Firstly, the performance of LISA on the OWASP Top 10 bugs for smart contracts~\cite{OWASP} is evaluated, determining the performance of LISA on the most common vulnerabilities. Secondly, several auditing projects are being conducted. used to evaluate the performance of the AI auditing tools. Finally, a real-world project is used to find out the pros and cons of current AI auditing tools, including LISA.

\noindent \textbf{Tools Selection.} For AI based tools, we are using Nethermind~\cite{Nethermind}, Almanax~\cite{Almanax}, QuillShield~\cite{QuillShield}, SmartAuditor.AI~\cite{SmartAuditor}, BevorAI~\cite{BevorAI}. For static analysis tools, we use the most popular Slither~\cite{slither}.

\subsection{Evaluation on OWASP Top 10}

\begin{table}[]
\caption{Performance on OWASP Top 10}
\label{tab:eval1}
\resizebox{\linewidth}{!}{
\begin{tabular}{llccccccc}
\toprule
OWASP ID  & Vulnerability Type              & LISA & Nethermind & Almanax & QuillShield & SmartAuditor.AI* & BevorAI & Slither \\
\midrule
SC01:2025 & Access Control Vulnerabilities  & \checkmark & \checkmark & \checkmark & x & NA & \checkmark & x  \\
SC02:2025 & Price Oracle Manipulation       & \checkmark & \checkmark & \checkmark & x & NA & \checkmark & x  \\
SC03:2025 & Logic Errors                    & \checkmark & \checkmark & \checkmark & x & NA & \checkmark & x  \\
SC04:2025 & Lack of Input Validation        & \checkmark & \checkmark & \checkmark & x & NA & \checkmark & x  \\
SC05:2025 & Reentrancy Attacks              & \checkmark & \checkmark & \checkmark & x & NA & \checkmark & \checkmark  \\
SC06:2025 & Unchecked External Calls        & \checkmark & \checkmark & \checkmark & x & NA & \checkmark & \checkmark  \\
SC07:2025 & Flash Loan Attacks              & x          & x          & x          & x & NA & x          & x  \\
SC08:2025 & Integer Overflow and Underflow  & \checkmark & \checkmark & \checkmark & x & NA & \checkmark & \checkmark  \\
SC09:2025 & Insecure Randomness             & \checkmark & \checkmark & \checkmark & x & NA & \checkmark & \checkmark  \\
SC10:2025 & Denial of Service (DoS) Attacks & \checkmark & \checkmark & \checkmark & x & NA & \checkmark & x  \\
\bottomrule
\multicolumn{9}{l}{*SmartAuditor.AI only support on-chain address as an input.}
\end{tabular}
}
\end{table}

In the first evaluation task, we collect cases from OWASP Top 10 2025 vulnerability types from OWASP's website, and the vulnerable cases of each vulnerability types. Comments related to vulnerability types are removed to avoid data leakage.

Table \ref{tab:eval1} presents the performance of LISA alongside several comparison tools when tested against the OWASP Top 10 (2025) vulnerability types for smart contracts. LISA matches or exceeds its peers in coverage in all ten categories. 
Since all the code segments for OWASP Top 10 are very simple, other AI auditing tools, including Nethermind, Almanax and BevorAI are able to detect most of them, except for the flash loan case, which is due to the inter-contract interactions not captured by any of the tools.
In contrast, QuillShield does not detect these vulnerabilities.
The static analysis based tool, Slither, can cover reentrancy, unchecked external calls, integer overflow and underflow, and insecure randomness. 
These vulnerability types can be described by static analysis rules while the other 6 types need semantic understanding, which is impossible for pure static analysis methods.

\subsection{Real-world Project Auditing}

\begin{table}[]
\caption{Performance on 5 real-world auditing projects}
\label{tab:eval2}
\resizebox{\linewidth}{!}{
\begin{tabular}{llccccccc}
\toprule
Project                                                                                                                                & Type                 & LISA & Nethermind & Almanax & QuillShield & SmartAuditorAI & BevorAI & Slither \\
\midrule
XLaunch              & Accounting errors    & \checkmark    & \checkmark          & x                                    & x                                        & NA                                          & x   & x                                 \\
QAMarketplace        & Accounting errors    & \checkmark    & x          & x                                    & x                                        & NA                                          & x       & x                             \\
ProofOfContribution  & Accounting errors    & \checkmark    & x          & x                                    & x                                        & NA                                          & x         & x                           \\
PauserRegistry       & State Inconsistency  & \checkmark    & \checkmark          & x                                    & x                                        & NA                                          & x            & x                        \\
SignPuff             & Missing state update & \checkmark    & \checkmark          & \checkmark                                    & x                                        & NA                                          & x           & x                        \\
\bottomrule
\end{tabular}
}
\end{table}

In the second evaluation, we selected five auditing contest projects, each containing high-severity vulnerabilities, to compare LISA’s detection capability against several other tools. These projects involved various complex logic bugs of vulnerability, such as accounting errors, state inconsistency, and missing state updates. As shown in Table \ref{tab:eval2}, LISA successfully detects all the targeted high-severity vulnerabilities across these five projects, while many competing tools fail to detect the same issues in several cases.

For example, in the XLaunch project, which contains accounting errors, LISA and Nethermind both identify the critical bug, whereas Almanax, QuillShield, BevorAI do not. In QAMarketplace and ProofOfContribution, LISA detects accounting errors that none of the other tools, except LISA, manage to catch. In PauserRegistry, which exhibits state inconsistency, Nethermind detects the issue along with LISA, but most other tools again fail. The SignPuff project has a missing state update vulnerability, which is detected by LISA, Nethermind, and Almanax but not by QuillShield or BevorAI.

These results indicate that LISA is particularly strong in catching more subtle and project-specific issues, errors in accounting logic, missing updates to state variables, and inconsistent state transitions, which are often missed by static pattern tools or simpler analysis pipelines. Complementing this, empirical studies (e.g., “Understanding Inconsistent State Update Vulnerabilities in Smart Contracts”) show that such state inconsistencies account for approximately 18\% of reported bugs in Code4rena audits. 
This supports our finding that logic and state‐update related issues are common in real audit contexts and require detection approaches that go beyond pattern matching.

While this experiment focuses on a small number of projects, the consistency of LISA’s success in these harder-to-detect cases suggests that its architecture (with the knowledge base, specialized agents, and fallback general agent) is effective in handling complexity and diversity of vulnerabilities in real human audits.

\subsection{Detailed Bug Coverage on a Selected Project}

The third evaluation was conducted on Size Meta Vault, which is an ERC4626 vault, and the evaluation result is analyzed by third-party auditors~\cite{AIAuditingBenchmark}.
As required by the auditor, the names of the tools are anonymized.
There are a total of 12 high and median level bugs in the given contract, and the evaluation result is shown in Table~\ref{tab:eval3}.

The table shows that LISA detects several of the medium severity bugs (notably issues M01, M03, M06, M07) but fails to detect the high severity issues (H01, H02, H03) in this particular version. The leading human auditor (Obsidian) catches many of these high severity issues, achieving full detection on H01–H03, but misses several of the medium ones. Tool 1 and other tools vary in their ability, often detecting some medium severity issues but failing on others and none matching the full breadth that Obsidian achieves.

These results illustrate both strengths and gaps of LISA. On the strength side, LISA's performance on medium-severity bugs is competitive: In practical auditing settings, detecting medium-severity issues can strongly improve a project’s security posture by addressing frequently occurring logic or state inconsistency bugs. On the gap side, the inability in this audit to detect high severity issues highlights limitations: either the patterns required to catch those bugs are not yet present in LISA’s Knowledge Base, or the analytical depth and agent specialization required to reason about certain high severity issues (for example, complex inter-strategy interactions, upgradeability or modular dependency risks) are still under development.

For H01 and H03, which are bugs which cannot be found by any of the existing tools, they involve the inconsistency between the developer's intention and the implementation.
H01 is an incorrect calculation of \texttt{totalAssets()} in the meta-valley that enables the stealing of depositor funds. 
When attackers deposit directly into the underlying strategies, the price of the assets in the vault will increase directly.
Similarly, H03 is about the \texttt{removeStrategies} function can be sandwiched to manipulate the price per share and steal deposited assets.
Current AI-based audit tools fail to capture the inter-contract design pattern and guess the intention of human developers.
In addition, LISA is the only AI-powered tool to detect M03, which is a sandwich attack with the removal of strategies, showing the potential generalizability of LISA.

Moreover, this audit also reveals trade-offs between detection coverage and detection confidence. Some of the high-severity issues may require deeper semantic reasoning or knowledge of intended protocol invariants (which are often external to the code itself) to confirm. In those cases, even specialized detection agents can miss them or produce lower confidence, leading to them being dropped or flagged for manual review.

In summary, the Size Meta Vault audit shows that while LISA can detect a meaningful subset of real audit bugs, particularly medium-severity ones, there remains important work to expand its coverage for high-severity, difficult-to-spot vulnerabilities, possibly by extending specialized agent definitions, incorporating richer project invariants, or enriching the Knowledge Base with more complex past high-severity audit cases.

\begin{table}[]
\caption{Performance on Size Meta Vault v0.0.1 Audit}
\label{tab:eval3}
\resizebox{\linewidth}{!}{
\begin{tabular}{lllllllllllll}
\toprule
Issue           & H01 & H02 & H03 & M01 & M02 & M03 & M04 & M05 & M06 & M07 & M08 & M09 \\
\midrule
Obsidian Audits & \checkmark  & \checkmark  & \checkmark  & \checkmark  & \checkmark  & \checkmark  & \checkmark  & x   & x   & x   & x   & \checkmark  \\
Tool 1      & x   & x   & x   & \checkmark  & \checkmark  & x   & \checkmark  & x   & \checkmark  & \checkmark  & \checkmark  & \checkmark  \\
Tool 2      & x   & \checkmark  & x   & \checkmark  & x   & x   & \checkmark  & \checkmark  & \checkmark  & x   & x   & x   \\
Tool 3      & x   & \checkmark  & x   & \checkmark  & x   & x   & \checkmark  & \checkmark  & \checkmark  & x   & x   & x   \\
LISA            & x   & x   & x   & \checkmark  & x   & \checkmark  & x   & x   & \checkmark  & \checkmark  & x   & x   \\
Tool 5      & x   & \checkmark  & x   & x   & x   & x   & x   & \checkmark  & \checkmark  & x   & x   & x   \\
Tool 6      & x   & x   & x   & x   & x   & x   & \checkmark  & x   & x   & x   & x   & x   \\
Tool 7      & x   & x   & x   & x   & x   & x   & x   & x   & x   & x   & x   & x   \\
Tool 8      & x   & x   & x   & x   & x   & x   & x   & x   & x   & x   & x   & x   \\
Tool 9      & x   & x   & x   & x   & x   & x   & x   & x   & x   & x   & x   & x   \\
Tool 10     & x   & x   & x   & x   & x   & x   & x   & x   & x   & x   & x   & x   \\
\bottomrule
\end{tabular}
}
\end{table}

\subsection{Detection of Real-world Attack Events}

\begin{table}[]
\caption{Recent attack events can be captured by LISA.}
\label{tab:eval4}
\resizebox{\linewidth}{!}{
\begin{tabular}{llll}
\toprule
Date       & Project             & Loss   & Description                                                                                                   \\
\midrule
2025-09-18 & Now Gold Protocol   & \$2M   & Contract steals liquidity from the Uniswap pair by transferring tokens directly from the pair's balance.      \\
2025-08-25 & ShibaSwap           & \$27K  & Lack of slippage protection in swaps leading to front-running and value loss.                                 \\
2025-08-14 & Grizzifi            & \$61K  & Team milestones based on total investments (including withdrawn) instead of active investments.               \\
2025-07-29 & SuperRare           & \$730K & Incorrect access control in `updateMerkleRoot` allows unauthorized users to update the merkle root.           \\
2025-07-15 & Arcadia Finance     & \$3.5M & Arbitrary external call in \_swapViaRouter allows initiator to perform unauthorized swaps                     \\
2025-07-11 & BankrollNetworkLife & \$113K & Profit calculation in distribute() does not cap to available balances, leading to over-issuance of dividends. \\
2025-07-05 & RANTToken           & \$203K & Incorrect burn amount calculation leading to price manipulation.                                              \\
2025-06-25 & Silo Finance        & \$550K & Profit calculation in distribute() does not cap to available balances, leading to over-issuance of dividends. \\
2025-06-19 & BankrollNetwork     & \$65K  & Incorrect payout adjustment in sell function leads to dividend miscalculations                               \\
\bottomrule
\end{tabular}
}
\end{table}

Beyond controlled benchmarks and contest datasets, we further demonstrate the applicability of LISA by analyzing a set of recent real-world exploits in decentralized finance (DeFi). Table~\ref{tab:eval4} summarizes incidents in which vulnerable projects suffered substantial financial losses, which could have been prevented had issues been identified during pre-deployment auditing with LISA. In all of these cases, the cumulative losses exceeded USD 7.2 million.

The incidents span a diverse range of vulnerability types. In the New Gold Protocol (September 2025), more than 2 million USD was lost due to a liquidity draining exploit that transferred tokens directly from the balance of the Uniswap pair. In Arcadia Finance (July 2025), a faulty external call in \_swapViaRouter enabled unauthorized swaps, leading to losses of more than USD 3.5 million. Similarly, SuperRare (July 2025) and RANTToken (July 2025) suffered USD 730,000 and USD 203,000 in damages, respectively, caused by incorrect access control in the Merkle root update and an erroneous burn-amount calculation that enabled price manipulation. Smaller but still significant losses were observed in Silo Finance (USD 550,000), BankrollNetworkLife (USD 113,000) and BankrollNetwork (USD 65,000), where flawed profit and payout logic resulted in overissuance or miscalculated dividends. Even relatively modest incidents, such as ShibaSwap (USD 27,000) and Grizzifi (USD 61,000), highlight the cost of overlooking logic bugs such as missing slippage protection or misuse of milestone-based investments.

These examples illustrate two important points. First, many high-impact vulnerabilities in DeFi are not simple pattern-matching errors but rather logic flaws and invariant violations that require semantic reasoning to detect. Second, the economic impact of missed vulnerabilities is disproportionate: Small oversights in logic can escalate into multi-million-dollar losses. By combining a knowledge-driven detection framework, specialized agents for known vulnerability classes, and invariant-aware reasoning, LISA is positioned to mitigate such incidents before deployment. As the data show, the adoption of LISA in these real-world cases could have saved more than USD 7.2 million in user and protocol funds, underscoring its practical importance for the Web3 ecosystem.

\section{Limitations}

While LISA and some of the AI auditing tools offer strong performance in many real-world auditing scenarios, and improves over existing tools on both vulnerability type coverage and accuracy, there remain important limitations shared across LISA and current AI-based auditing solutions. Understanding these limits is essential for interpreting results correctly, planning improvements, and deploying such tools in production.

First, coverage gaps in vulnerability types remain a key issue. Many auditing contests and real-world codebases have already had common, simple vulnerabilities (e.g. basic reentrancy, integer overflow/underflow, trivial access control misconfigurations, missing input validation) detected and often patched before further analysis. Because such simple bugs are underrepresented in post-audit contest code or high maturity contracts, tools like LISA may be less tested (and less strong) on those trivial cases in some datasets. Additionally, certain high-severity flaws, especially those involving complex inter-contract interactions, subtle economic invariants, upgradeability patterns, or protocol-level invariants, are often not well captured in knowledge bases or pattern heuristics, leading to misses or lower confidence.

Second, dependence on historical data / past audit reports imposes limitations. LISA’s Knowledge Base approach learns patterns, heuristics, and examples from past audits, which improves generalization to similar projects. But this also means that when novel vulnerability types emerge, especially ones not well represented in historical audits, detection may lag. Variant coding styles, new language constructs, evolving design patterns of DeFi protocols or smart contracts, or new attack vectors that were not previously seen can challenge detection agents that rely on past examples.

Third, false positives and false negatives, especially in logic or semantic vulnerabilities, are still a challenge. AI/LLM-based tools may misinterpret intent, produce spurious findings, or miss bugs whose manifestation depends on dynamic execution, hidden state transitions, or external inputs/data. Static analysis, often used in combination with AI, can flag code that appears vulnerable but in practice is defended by guards or context that the analysis doesn’t fully account for. On the flip side, dynamic or contextual vulnerabilities may be missed if the tool lacks sufficient execution model, context, or inter-contract dependency information.
In this technical report, only the coverage of vulnerability is considered, but in real world scenarios, human auditors will also be aware of the precision of the audits.

Fourth, explainability, confidence, and traceability are often weak. It is one thing for a tool to flag “this function might have state inconsistency” or “this oracle call could be manipulated,” but stakeholders (developers, auditors, risk management) need to understand why the detection was made. Which past knowledge entry or pattern was matched? What code paths are involved? What invariants are proposed to be violated? Without clear explanation or confidence scores, flagged issues may require additional manual effort to triage, reducing efficiency gains. All existing tools fail to give accurate and convincing reasons for bugs found, which may confuse users.

Fifth, scalability, performance, and runtime constraints matter. Some AI-driven tools incur high computational cost or take long time on large codebases. Large language model based methods may require substantial memory or compute resources, which might make them impractical for rapidly iterated audits or continuous integration pipelines. In addition, contract size, modular dependencies, and external interactions can markedly increase analysis complexity and runtime.

Finally, project-level invariants or protocol semantics are often external to the code and may not be fully available. Some vulnerabilities can be detected only when knowing invariant policies (e.g. token supply invariants, oracle update schedules, cross-contract trust boundaries) which are not encoded in code or are not explicit in project documentation. Without them, detection is necessarily approximate, with the risk of missing important flaws or assigning wrong severity.

\section{Conclusion}

This report has introduced LISA, an agentic smart contract vulnerability detection framework that unifies rule-based and logic-based methods, enriched by a Knowledge Base built from historical audit reports. Across multiple evaluations, including the OWASP Top 10 benchmark, selected real-world auditing contest projects, and the Size Meta Vault v0.0.1 audit, LISA demonstrates strong vulnerability type coverage (especially for logic errors, state inconsistencies, and medium-severity bugs), as well as the ability to generalize to previously unseen code bases without requiring model fine-tuning.

The LISA architecture, which combines a rich and evolving knowledge base, a Scheduler that dynamically deploys specialized and general agents, and a careful result merging process (factual check, cross-validation against historical patterns, and project-level invariants), contributes to higher detection accuracy and better practical usability compared to many existing tools. In contexts where simple rule‐based vulnerabilities have already been addressed upstream (e.g., code already audited or prescanned by static tools), LISA provides added value by focusing on subtler logic bugs and contract semantics that are often missed.

However, the evaluations also highlight limitations: high-severity issues in complex contracts sometimes remain undetected; coverage of certain novel vulnerability types is still incomplete; false negative and false positive risks persist (especially when project invariants or external context are not fully specified). Additionally, performance (in terms of runtime, resource use) and explainability remain critical areas for refinement, especially for deployment in production audit pipelines.

Looking ahead, there are several promising directions to further enhance LISA. Expanding the Knowledge Base to include more high-severity and less common vulnerabilities will help reduce coverage gaps. Developing more specialized agents for domain-specific contract patterns (e.g., DeFi vaults, cross-chain bridges, upgradable contracts) will improve detection of complex bugs. Improving invariant specification tools so that project-level invariants (business logic, economic invariants) can be more explicitly captured and used in merging will strengthen both precision and relevance. Finally, optimizing for scalability and integrating LISA into continuous integration / deployment workflows will help bring its benefits into the hands of developers and auditors on a routine basis.

In sum, LISA represents a meaningful advance in smart contract auditing: By combining learning from past audits, agent orchestration, and logic-aware analysis, it narrows the gap between automated tools and expert auditors. Although it is not yet perfect, its performance in real-world settings shows that it is a viable component of a secure smart contract development lifecycle. Future work will aim to make it more comprehensive, more explainable, and more broadly applicable.

\bibliographystyle{unsrt}
\bibliography{ref}


\end{document}